\documentclass{ws-ijmpd2}

\usepackage[super,compress]{cite}
\usepackage{color}
\usepackage{nccmath}
\usepackage{moresize}
\usepackage{enumerate}
\usepackage[caption=false]{subfig}
\usepackage{stackengine}
\newcommand{\be}{\begin{equation}}
\newcommand{\ee}{\end{equation}}
\newcommand{\bea}{\begin{eqnarray}}
\newcommand{\eea}{\end{eqnarray}}
\newcommand{\bse}{\begin{subequations}}
\newcommand{\ese}{\end{subequations}}
\newcommand{\bce}{\begin{center}}
\newcommand{\ece}{\end{center}}
\newcommand{\bfg}{\begin{figure}}
\newcommand{\efg}{\end{figure}}
\newcommand{\bit}{\begin{itemize}}
\newcommand{\eit}{\end{itemize}}
\newcommand{\bed}{\begin{description}}
\newcommand{\eed}{\end{description}}
\newcommand{\ben}{\begin{enumerate}}
\newcommand{\een}{\end{enumerate}}
\newcommand{\nn}{\nonumber}

\newcommand{\fr}{\frac}
\newcommand{\sq}{\sqrt}
\newcommand{\no}{\noindent}

\def\le {\left}
\def\ri {\right}
\def\r  {\rho}

\newcommand{\cR}{\mathcal R}
\newcommand{\cS}{\mathcal S}

\newcommand{\vx}{\vec{\pmb x}}

\newcommand{\bdm}{\begin{displaymath}}
\newcommand{\edm}{\end{displaymath}}

\begin{document}

\markboth{Saurya Das and Sourav Sur}
{Emergent gravity at all scales}

%
%

\title{\Large Emergent gravity at all scales}

\author{SAURYA DAS}

\address{\it Theoretical Physics Group and Quantum Alberta\\
Department of Physics and Astronomy, University of Lethbridge\\
4401 University Drive, Lethbridge, Alberta T1K 3M4, Canada\\
saurya.das@uleth.ca; saurya.das@gmail.com}

\author{SOURAV SUR}

\address{\it Department of Physics and Astrophysics, University of Delhi\\
New Delhi - 110 007, India\\
sourav.sur@gmail.com; sourav@physics.du.ac.in}

\maketitle

\begin{history}
\received{}
\revised{}
\end{history}


%


%
%



\begin{abstract}
It has recently been shown that any observed potential can in principle 
be generated via quantum mechanics using a suitable wavefunction.
In this work, we consider the concrete example of the gravitational
potential experienced by a test particle at length scales spanning from the 
planetary to the cosmological, and determine the wavefunction that 
would produce it as its quantum potential. In other words, the observed gravitational interaction at all length scales can be 
generated by an underlying wavefunction. 
We discuss implications of our result. 
\end{abstract}

\maketitle

\vspace{10pt}
\noindent 
Can all perceived potentials in nature have a simple quantum mechanical origin? 
We asked this question recently, and showed that a quantum particle sees the sum 
of the background classical potential and a wavefunction dependent quantum 
potential, and there is no way of separating the two in any experiment
\cite{qp1,qp2}.
In other words, potentials around us can be attributed to quantum mechanics, 
either partially or wholly, and in the case of the latter, the background 
spacetime is flat and with no classical interactions, but `filled' with suitable wavefunctions. 

We would test this idea in this paper in the context of gravity, considering in 
particular, 
%
the effective gravitational potential $V(r)$, on a test particle of mass $m$,
to be given by
%
\bea 
V (r) \,=\, \le\{ \begin{aligned}
&V_1 = - \dfrac{k_1} r \,; ~ &r\leq r_1 \,,
\\
&V_2 = k_2 \,\ln \le(\dfrac r {r_0}\ri) ; ~ &r_1 < r\leq r_2 \,, 
\\
&V_3 = - \dfrac 1 2 \,k_3\, r^2 \,; ~
& r_2 < r < \infty \,, 
\label{pot}
\end{aligned} 
\ri. 
\eea 
%
%
%
%
%
%
where $r_1$ and $r_2$ are respectively the galactic and the cosmological length scales, $k_1, k_2, k_3$ and $r_0$ are dimensionful constants with $r_0$ being arbitrary. 
Note that this effective potential can explain observations at all scales, 
without 
making any specific allusion to dark matter or dark energy. A similar potential 
stretching across length scales has also been proposed in the literature
%
\cite{sivaram,haranas}.

Now, in an earlier set of works, it was shown that 
practically any potential experienced by a particle
can in principle 
have its origin in quantum mechanics in terms of its wavefunction $\Psi$. 
The latter gives rise to its own `quantum potential', the particle sees the sum of the background classical and the new wavefunction dependent quantum potential, and 
there is no way of distinguishing between the two 
\cite{qp1,qp2}.
In other words, in principle the potential (\ref{pot}) may entirely be the quantum potential of a suitable wavefunction of the particle. Before we compute that wavefunction, we first present a brief review of the derivation of the quantum potential. 

Consider the time-dependent Schr\"odinger equation for a point particle of mass $m$
\begin{equation}
-\frac{\hbar^2}{2m} \nabla^2 \Psi + V \Psi = i\hbar\frac{\partial \Psi}{\partial t}.
\label{cos2}
\end{equation}
We write the wave function as
\begin{equation}
\Psi (\vec r,t) =\, {\cal R} (\vec r,t)\, e^{i S (\vec r,t)}~,
\end{equation}
where ${\cal R} (\vec r,t)$ and $S (\vec r,t)$ are real functions. Substituting this in Eq.(\ref{cos2}) and extracting the real and imaginary parts, we obtain 
\begin{equation}
-\,\frac{\partial S}{\partial t} =\, \frac{(\nabla S)^2}{2m}+ V(r)
-\frac{\hbar^2}{2m}
\frac{\nabla^2 {\cal R}}{{\cal R}}~,
\label{cos3}
\end{equation}
and the continuity equation
\begin{equation}
\frac{\partial \rho}{\partial t}+\nabla.(\rho \frac{\nabla S}{m})=0~,
\end{equation}
where $\rho={\cal R}^2$.
In the absence of the last term in Eq.(\ref{cos3}), namely
\begin{equation}
V_Q=-\frac{\hbar^2}{2 m} 
\frac{\nabla^2 {\cal R}}{{\cal R}}
\label{cos4}
\end{equation}
we are left with the classical Hamilton-Jacobi equation, 
giving rise to 
the standard Newtonian trajectories for the constituent particles. 
However, if we define the `velocity field' as 
\be 
v (\vec r,t) \equiv\, \fr{\hbar} m \, \vec\nabla S~,
\ee 
Eq.(\ref{cos3}) can simply be written as Newton's second law of motion with the standard classical potential (or force) augmented by the quantum potential (and a corresponding wavefunction dependent force) as follows
\cite{bohm1,bohm2,holland}
\bea 
&& m \, \fr{d\vec v}{dt} =\, - \, \vec\nabla \le(V +\, V_Q\ri)~. 
\label{Newt-eq}
\eea 
Evolution via $(V+V_Q)$ yields the so-called 
`Bohmian trajectories' or 
`quantal trajectories', which can be thought of as classical trajectories plus the ${\cal O}(\hbar^2)$ quantum corrections.
%
As expected, these become important at short distances
(when quantum effects are expected to dominate) and indeed reproduce
the results of all quantum experiments, such as the interference patterns
in a double slit experiment. It is implicit of course that no intermediate measurements are done on the quantal trajectories until the particle reaches its desired destination and standard quantum mechanical measurements are made at that point. 
The Bohmian interpretation of quantum mechanics,
albeit not the most popular one, is perfectly equivalent to the standard ones,
offers a novel and useful picture in many situations,
such as the current one and 
elsewhere \cite{barth,mollai}. 
In other words, all predictions from the (Bohmian) approach are producible from the 
standard formalism of quantum mechanics, albeit in a slightly 
different way. The generalization of the above method to relativistic scenarios raises some interesting interpretational issues regarding the quantal trajectories, but the procedure as such is straightforward and once again yields identical predictions as from the conventional approach.  

With the above, we now return to
the question as to 
whether it is plausible to have a wave-function $\Psi$ that
can give rise to the above potential (\ref{pot}) as a {\em quantum potential} 
%
%
%
%
%
over the entire range of $r$. Or, in other words, can one have the following?
\bea 
\Psi \,=\, \le\{ \begin{aligned}
&\Psi_1 \rightarrow V_Q = V_1 \,; ~ &r\leq r_1 \,,
\\
&\Psi_2 \rightarrow V_Q = V_2 \,; ~ &r_1 < r\leq r_2 \,, 
\\
&\Psi_3 \rightarrow V_Q =  V_3 \,; ~
& r_2 < r < \infty \,.
\label{qpot}
\end{aligned} 
\ri. 
\eea 
%
%
%
%
%
%
%
%
If so, then one can indeed say that the wavefunction `induces' the observed 
potential and that gravity `emerges' from an underlying quantum theory in flat 
spacetime.

Re-writing Eq.(\ref{cos4}) as 
\begin{eqnarray}
&& {\nabla^2 \cR}  + \frac{2m\,V_Q}{\hbar^2}\,{\cal R} = 0~,
\label{qp2}
\end{eqnarray}
a number of important points may be noted:
\ben[(i)]
\item $V_Q$ is a local quantity. 
\item $V_Q$ depends on the amplitude $\cR$, but not on the phase $\cS$, of 
the wavefunction $\Psi$.  
\item $V_Q$ does not depend on the overall normalization of the wavefunction
$\Psi$. Thus, for example, if $\Psi$ is suddenly magnified by an arbitrary 
factor in a region of space, there will be no effect on the quantum potential. 
\item Despite the similarity, Eq.(\ref{qp2}) should not be confused with the 
{\it free} time-independent Schr\"odinger equation for stationary states, 
namely,
\be
\nabla^2\,\Psi_s + \fr{2mE}{\hbar^2}\,\Psi_s 
= 0 \,,
\ee
which does not involve any background classical potential. Here, $E$ is a 
constant, $\Psi_s$ is a stationary state wavefunction, and the full 
wavefunction $\Psi$ can be considered as a superposition of a requisite 
number of $\Psi_s$. 
\een 
It follows from the point (ii) that the phase $\cS$ remains undetermined. 

Next, we solve Eq.(\ref{qp2}) by substituting for $V_Q$ in the three regions, 
given by Eq.(\ref{pot}). 
Since the potential 
is spherically symmetric, we assume, for simplicity, the wavefunction to be 
spherically symmetric as well, i.e., $\cR = \cR (r)$. We also assume the phase 
of the wavefunction $\cS=0$. Eq.(\ref{qp2}) then simplifies to
\be
\fr 1 {r^2} \, \fr d {dr} \le(r^2\, \fr{d\cR}{dr}\ri) + \fr{2m\,V_Q}{\hbar^2}
\,{\cal R} =\, 0 \,.
\label{qp4}
\ee
Solutions which are not spherically symmetric and for which the phase $\cS$ is 
non-vanishing will be investigated in a future work. 

For the potential given by Eq.(\ref{pot}),
the above Eq.(\ref{qp4}) translates to the following set of equations
\bea
&&\hspace{-25pt} \fr 1 {r^2} \Big[r^2\, \cR'(r)\Big]'\!
- \fr 1 {\ell_1\,r}\,\cR(r) =\, 0\,; ~~\quad\quad\quad r \leq r_1 \,,
\label{qp5} \\
%
&&\hspace{-25pt} \fr 1 {r^2} \Big[r^2\, \cR'(r)\Big]'\!
+ \fr 1 {\ell_2^2} \ln \!\le(\!\fr r {r_0}\!\ri) \cR(r) = 0\,; \quad
r_1 < r \leq r_2 \,, 
\label{qp6} \\
%
&&\hspace{-25pt} \fr 1 {r^2} \Big[r^2\, \cR'(r)\Big]'\!
- \fr{r^2}{\ell_3^4}\,\cR(r) =\, 0\,; \qquad\quad\quad r_2 < r < \infty \,,
\label{qp7}
\eea
where the prime $\{'\} \equiv d/dr \,$, and 
%
\bea
&& \fr 1 {\ell_1} =\, \fr{2m\, k_1}{\hbar^2} \,,\\
&& \fr 1 {\ell_2^2} =\, \fr{2m\,k_2}{\hbar^2} \,, \\
&& \fr 1 {\ell_3^4} =\, \fr{m k_3}{\hbar^2} \,.
\eea
Note that $\ell_1,\ell_2,\ell_3$ all have dimensions of length. 

The general solutions in the three regions can be written symbolically as
\bea
\Psi_1(r) &\equiv& \cR_1(r) =\, c_1\, f_1(r) \,; \qquad\qquad\qquad 
r\leq r_1 \,, 
\label{Psi1} \\
\Psi_2(r) &\equiv& \cR_2(r) =\, c_2\, f_2(r) +\, d_2\, g_2(r) \,; ~\quad
r_1 \leq r\leq r_2 \,,
\label{Psi2} \\
\Psi_3(r) &\equiv& \cR_3(r) =\,
d_3\, g_3(r) \,; \qquad\qquad\qquad
r_2 < r < \infty \,, \label{Psi3}
\eea
where $c_1, c_2, d_2$ and $d_3$ are constant coefficients of  
the respective linearly independent solutions of relevance, denoted by 
$c_1,c_2,d_2$ and $d_3$.
Note that in either of the Eqs.(\ref{Psi1}) and (\ref{Psi3}), only one of 
the two linearly independent solutions is of relevance, namely $f_1(r)$ and
$g_3(r)$, which remain finite respectively at $r = 0$ and as $r \rightarrow 
\infty$.
%
Eq.(\ref{Psi2}), on the other hand, involves both the linearly independent 
solutions, $f_2(r)$ and $g_2(r)$. 
While the exact forms of $f_1(r)$ and $g_3(r)$ can be derived from the 
respective Eqs.(\ref{Psi1}) and (\ref{Psi3}) in a straightforward way,
solving Eq.(\ref{qp5}) to obtain $f_2(r)$ and $g_2(r)$ is not quite simple,
because of the presence of the logarithmic term. 
We therefore solve Eq.(\ref{qp5}) by approximating the logarithm to a 
polynomial, i.e., 
\be 
\ln \le(\fr r {r_0}\ri) =\, \ln \le(1 +\, \fr{r-r_0}{r_0}\ri) =\, 
\fr{r-r_0}{r_0}  
+\, \dots \,,
\ee
and retaining only the leading order term, after choosing the arbitrary 
constant $r_0$ such that $r \ll r_0$. One can of course obtain the 
solution to any desired accuracy by retaining more and more terms, or 
by solving numerically.
We have, finally
\bea
&&\hspace{-25pt} f_1 =\, \sq{\fr{\ell_1} r}\, I_1 
\le(2\, \sq{\fr r {\ell_1}}\ri) \,, \\
&&\hspace{-25pt} f_2 =\, \fr 1 r \, Ai \le((-1)^{1/3}\,
\le[\fr{r_0}{\ell_2}\ri]^{2/3}
\le[\fr r {r_0} - 1\ri] 
\ri) +\, \text{c.c.} \,, \\
&&\hspace{-25pt} g_2 =\, \fr 1 r \, Bi \le((-1)^{1/3}\,
\le[\fr{r_0}{\ell_2}\ri]^{2/3}
\le[\fr r {r_0} - 1\ri]
\ri) +\, \text{c.c.} \,, \\
%
%
&&\hspace{-25pt} g_3 =\, \fr{e^{-r^2/\ell_3^2}} r \,
H_{-1/2} \le(\fr r {\ell_3}\ri) \,, 
\eea
where $I_n, Ai, Bi, H_n$ are the Bessel function, Airy $Ai$ and $Bi$
functions and Hermite polynomials, respectively, and c.c. denotes the
complex conjugation.

Next, we impose the following matching conditions for the wavefunction 
$\Psi(r)$ and its derivative $\Psi'(r)$ at two points $r=r_1$ and $r=r_2$:
\bea
\Psi_I (r_1) &=& \Psi_{II} (r_1) \,, \\
\Psi_I' (r_1) &=& \Psi_{II}' (r_1) \,, \\
\Psi_{II} (r_2) &=& \Psi_{III} (r_2) \,, \\
\Psi_{II}' (r_2) &=& \Psi_{III}' (r_2) \,.
\eea
These translate, respectively, to the equations
\bea
&& c_1\, f_1(r_1) =\, c_2\, f_2(r_1) +\, d_2\, g_2(r_1) \,, \\
&& c_1\, f_1'(r_1) =\, c_2\, f_2'(r_1) +\, d_2\, g_2'(r_1) \,, \\
&& c_2\, f_2(r_2) +\, d_2\, g_2(r_2) =\, 
\, d_3\, g_3 (r_2) \,, \\
&& 
c_2\, f_2'(r_2) +\, d_2\, g_2'(r_2) =
\, d_3\, g_3'(r_2) \,.
\eea
Since there are four equations involving four constants, one can in
principle determine these constants. As mentioned before, the overall
normalization is not important for our purposes. However, if required, 
the wavefunction $\Psi$ can be easily normalized upon dividing it by 
$4\pi\,\int_0^\infty \, dr\,r^2 |\Psi|^2$.

We plot the quantum potential $V_Q(r)$ and the complete wavefunction 
$\Psi(r)$ in Figure 1, by matching the forces corresponding to the 
former's piecewise forms, $V_1$, $V_2$ and $V_3$, at the boundaries $r_1$ 
and $r_2$. Such a matching in fact helps to narrow down the parameter 
space, as it leads to two conditions involving the five parameters
$\{\ell_1,\ell_2,\ell_3,r_1,r_2\}$, meaning only three of them are 
independent. Then rescaling $r$ by $r_0$, we obtain the plots for
the fiducial settings $\, r_1 = 0.01 \,r_0, \, r_2 = 0.1 \, r_0\,$ 
and $r_0 = 10^4$ (for which the above approximation of the logarithmic
form of $V_2$ holds well). 
Further implications of the matching of the forces at the boundaries 
would be explored in a future publication. 

%
%
\begin{figure}[t]
    \centering
    \includegraphics[width=0.75\textwidth]{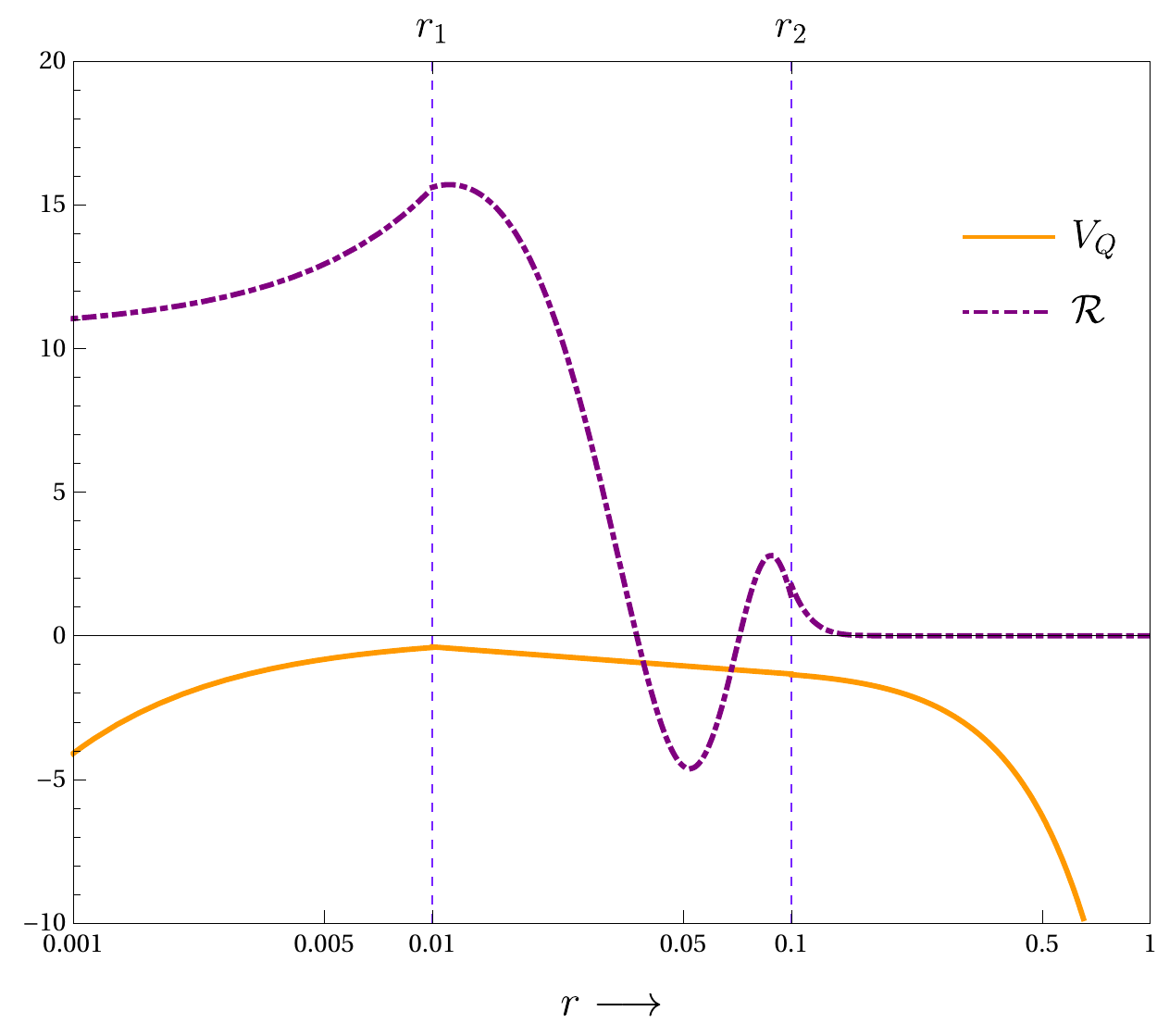}
    \caption{Plots of the quantum potential $V_Q$ in Eqs.(1)-(3) and its 
    source wavefunction $\cR$ for all $r$, which is made dimensionless
    by scaling with $r_{_0}$. A similar scaling is also done for $V_Q$ in 
    order to plot in the same graph, alongside $\cR$.
    }
    \label{fig:mesh1}
\end{figure}

To summarize, we have shown in this paper that the gravitational potential
which accounts for most observations, 
from the planetary to the cosmological scales, can be correctly reproduced 
by a suitable wavefunction via its quantum potential. The wavefunction is 
continuous and finite everywhere, and is normalizable. This supports the 
proposal that what appears as classical potential may in fact be the 
quantum potential of a suitable wavefunction in {\it flat and free space}. 
The symmetries of the background spacetime is that of the vacuum, namely 
the usual Poincar\'e symmetry and the potential and force is experienced 
by a particle purely due to its wavefunction. It is hard to think of a 
simpler physical picture, in which classical potentials and forces are 
emergent and completely producible from an underlying quantum picture. 
%
%
%
%
%

As to some future directions of work, one may stretch a comparison of the 
present approach with that in related works on gravitational and other 
potentials arising from an underlying quantum framework 
\cite{perelman,matone1,faraggi1,faraggi2,floyd,db1,grf20,db2,qp1,dss-qb,
ss-qb,faraggi3}.
One would need to study other wavefunctions compatible with the given 
potentials, for e.g., the ones with a non-trivial angular dependence, or 
phase. Moreover, one would need to generalize this formalism to make it 
manifestly (special) relativistic, as well as study the decoherence of the 
said wavefunction. 
However, given that there is no `environment' as such, it is unclear what 
would cause such a decoherence. In any case, a set of concrete predictions, testable in the astrophysical or cosmological scenario, would help strengthen 
our formalism here. We hope to report on these elsewhere.

\vspace{10pt}
\noindent 
{\bf Acknowledgment}

\no
This work was supported by the Natural Sciences and Engineering
Research Council of Canada.

\vspace{10pt}
\noindent 
{\bf Data availability statement}

\no
Data sharing not applicable to this article as no datasets were generated or analysed in this study.



\end{document}